\documentclass{llncs}

\usepackage{cite}

\usepackage{listings}
\usepackage[table]{xcolor}
\usepackage{xspace}
\usepackage{cite}
\usepackage[hyphens]{url}
\usepackage{flushend}
\usepackage{pgfplots, pgfplotstable}
\usetikzlibrary{patterns}
\usepackage{hyperref}
\usepackage{tikz}

\usepackage{paralist}
\usepackage{mathptmx}
\definecolor{mygreen}{rgb}{0,0.6,0}
\definecolor{mygray}{rgb}{0.5,0.5,0.5}
\definecolor{mymauve}{rgb}{0.58,0,0.82}

\pgfplotsset{compat=newest}

\newcommand{\toolname}{ROPocop\xspace}

\begin{document}

\title{ROPocop - Dynamic Mitigation of Code-Reuse Attacks}
%
%
%
%
%


\author{Andreas Follner, Eric Bodden}
\institute{Secure Software Engineering Group\\
       EC SPRIDE, Technische Universit\"at Darmstadt\\
\email{andreas.follner@cased.de}, \email{bodden@acm.org}
}
%
%

\maketitle
\vspace{-10pt}
\begin{abstract}

Control-flow attacks, usually achieved by exploiting a buffer-overflow vulnerability, have been a serious threat to system security for
over fifteen years.
Researchers have answered the threat with various mitigation techniques, but nevertheless, new exploits that
successfully bypass these technologies still appear on a regular basis.

In this paper, we propose \toolname, a novel approach for
detecting and preventing the execution of injected code and for mitigating
code-reuse attacks such as return-oriented programming (RoP). \toolname uses
dynamic binary instrumentation, requiring neither access to source code
nor debug symbols or changes to the operating system. It mitigates attacks by both
monitoring the program counter at potentially dangerous points and by detecting
suspicious program flows.

We have implemented \toolname for Windows x86 using PIN, a
dynamic program instrumentation framework from Intel. Benchmarks using the SPEC CPU2006 suite
show an average overhead of 2.4x, which is comparable to similar approaches,
which give weaker guarantees. Real-world applications show only an initially
noticeable input lag and no stutter. In our evaluation our tool successfully
detected all 11 of the latest real-world code-reuse exploits, with no false
alarms. Therefore, despite the overhead, it is a viable, temporary solution to secure critical systems against exploits if a vendor patch is not yet available.

\end{abstract}

\section{Introduction}
\label{Sec:Intro}
Attacks that aim at manipulating a program's control flow, often
through a buffer overflow vulnerability, are still one of the biggest threats to
software written in unsafe languages like
C or C++~\cite{bo_top_25}.
If successfully exploited, control-flow attacks can allow an adversary to execute arbitrary code. 
In the early 2000s, operating-system developers started adding mitigation techniques into their software. To this day, new techniques are added on a regular basis, however, while they make successful and reliable exploitation much more difficult, they can be bypassed.
Contests like, e.g., pwn2own~\cite{pwn2own} continuously show that current mitigation techniques are insufficient when it comes to protecting applications, and that more comprehensive methods are required. Currently, the most widely used attack technique, and an essential part of virtually every exploit, is RoP~\cite{rop}, where instead of injecting new code an attacker pieces together short code fragments, which already exist in memory. Recently proposed solutions against such attacks mostly built on CFI~\cite{cfi_for_cots, kbouncer, ccfir}, seemed effective, but have been shown to be bypassable~\cite{stitching_gadgets, size_matters}. 
Section~\ref{Sec:Motivation} elaborates on these issues in detail.

To battle current exploitation mechanisms we propose \toolname, a novel tool that
mitigates control-flow attacks for x86 Windows binaries using two novel
techniques, AntiCRA and DEP+. AntiCRA greatly reduces the risk of
successful code-reuse attacks by detecting an unusually high
rate of successive indirect branches during the execution of unusually short
basic blocks.
As different programs can exhibit very different behaviour in regards to that aspect, using the same threshold for every program is suboptimal. Therefore, \toolname comes with a learning mode, which runs ahead of time and determines appropriate thresholds, which can be adopted by the user. However, we do also provide default thresholds which work very well in practice and for a large selection of programs, as our evaluation shows.

Our second contribution, DEP+, implements a variant of a non-executable stack through dynamic
binary instrumentation. DEP+ assumes that all code
has to reside within an image. This is very similar to DEP~\cite{xp_sp2_dep}, however, DEP+ cannot be disabled
through API calls, thereby eliminating a large class of exploits that are based
on such calls. DEP+ enforces all memory to be non-executable, except for the
parts to which images are loaded. To this end, DEP+ monitors the loading and
unloading of images, checking after each indirect branch whether the program
counter points outside the known images.

We have implemented \toolname for Windows x86 using
PIN~\cite{Luk:2005:PBC:1065010.1065034}, a freely-available dynamic program
instrumentation framework from Intel.
\toolname requires no access to source code or root privileges, nor debug
symbols or changes to the operating system. Measurements using the artificial SPEC
CPU2006 suite show an average overhead of 2.4x. More importantly, experiments on real-world applications show only an initially
noticeable input lag (caused by the initial dynamic instrumentation) and no
stutter.
Our evaluation using 11 of the latest code-reuse exploits shows that
our tool successfully prevents all code-injection attacks and code-reuse attacks
from succeeding, even a highly sophisticated attack that relies solely on code reuse~\cite{reader_666}. Our envisioned usage of \toolname is to use it as a last line of defense against exploitation of critical systems, e.g., when a severe vulnerability has been discovered but no patch is available.

To summarize, this work makes the following original contributions:
\begin{compactitem}
	\item AntiCRA, a tunable heuristic detection of code-reuse-attacks like RoP and JoP,
	\item DEP+, a comparatively fast and robust implementation of a non-executable
	stack,
	\item \toolname, a dynamic instrumentation tool based on PIN
	which detects various kinds of control-flow attacks using the above
	techniques, and	
	\item an empirical evaluation showing that \toolname's mitigation approach is
	highly effective and shows tolerable runtime overheads.
\end{compactitem}

We make \toolname available online as open source, along with all our
experimental data: \url{https://sites.google.com/site/ropocopresearch/}


\section{Current Situation}
\label{Sec:Motivation}
Exploiting vulnerabilities with the goal to manipulate the program flow
was relatively trivial on Windows until the early 2000s, when Microsoft began
adapting mitigation techniques. In the
simplest cases, an attack widely known as \textit{stack smashing}~\cite{aleph}
could be used. Such an attack would leverage unbounded functions, such as
\texttt{strcpy}, to write beyond the allocated
memory of a buffer. Attackers could thus overwrite the function's stored
return address on the stack with an address that points to injected code,
which the program will execute after the next return.

To defend against such code injection attacks, Microsoft implemented
\textit{Data Execution Prevention} (DEP)~\cite{microsoft_dep}, which makes use
of a processor's NX (no execute) bit. DEP marks pages which contain data
as non-executable, causing a hardware-level exception if
execution from within such a page is attempted. This successfully prevents
attacks that attempt to execute injected code.

Nevertheless, attackers can bypass DEP in various ways. At present, the
most widely used technique is called return-oriented programming~\cite{rop_org}. When
utilizing RoP, an attacker does not inject any code but instead uses existing
code fragments (\textit{gadgets}), which all end with a return
instruction.
In other words, instead of injecting code, the attacker injects the addresses of
the gadgets he wants to execute. On x86, return works by popping an
address off the stack into the register EIP and then jumping to that
address, even if the stack is marked as non-executable. By crafting a stack
filled with a sequence of gadget addresses, the attacker can execute sequences
of gadgets, with the return instruction at the end of each gadget transferring
the program flow to the next gadget. Jump-oriented programming (JoP)~\cite{jop, jop2, jop_windows} is based on the
same basic concept as RoP, but uses \texttt{jmp} instructions to transfer control flow to the next
gadget.
In the following, we refer to both RoP and JoP attacks as
\emph{code-reuse attacks}.

The success of code-reuse attacks depends on the availability of
useful gadgets on the target platform and the complexity of the code the
attacker wants to run. In practice, however, most systems are vulnerable to
such code-reuse attacks. Furthermore, RoP attacks are relatively complex to stage, which is why most attacks
of this kind do not resort to pure RoP, but rather implement a
two-staged approach. The first stage uses RoP to call a Windows API
function like \texttt{VirtualProtect} (see below) which marks a certain memory
region as executable, effectively bypassing DEP. This is followed by the second stage, running
code previously injected into that memory region, which can then be executed
as normal.
%
%
Code-reuse attacks work reliably if the memory layout of an application is
highly deterministic because an attacker can hard-code the addresses of gadgets
directly into the exploit. To mitigate this, Microsoft introduced randomness in
the form of ASLR~\cite{microsoft_aslr}. ASLR randomizes the order in which images are
loaded into the virtual address space and adds pseudo-random offsets to their
base addresses. This makes it very difficult for an attacker to predict the
memory locations of the required gadgets on the target system.

\subsection{DEP Weaknesses}
Whether or not a program is protected by DEP depends on the compiler setting
the \verb$NX_COMPAT$ flag in the header of the program's main executable. This
flag may be left unset due to a number of reasons, including unsafe compiler
defaults and program incompatibility. Thus, the opt-in nature of DEP may render
its benefits void. Furthermore, DEP is either on or off for the whole program,
meaning that linking any one library with DEP disabled, be it statically or
dynamically, will cause DEP to be disabled for the entire process. If it is
enabled, an attacker can only bypass it through previously discussed
code-reuse attacks. Such attacks succeed without ever running
code from non-executable pages, which is why DEP cannot protect against them. The previously mentioned
two-staged attack, however, is worth explaining in more detail, since it allows injected code to be run,
effectively bypassing DEP. Such attacks work because Windows exposes
certain functions to change the permissions of a page (e.g.
\texttt{VirtualProtect}) or allocate new pages with specific permissions (e.g.
\texttt{VirtualAlloc}), which are passed as parameters. These APIs exist because
some programs generate code at runtime and therefore require memory that is both
writeable and executable.
Such program behaviour is strongly discouraged by Microsoft, though, and many
producers of software have adapted their programs to be compatible with DEP.

\subsection{ASLR Weaknesses}
Like DEP, ASLR is not necessarily enabled, depending on the flag
\texttt{DYNAMIC\char`_BASE}. However, unlike DEP, it can be localized to specific
modules, with the operating system being able to handle processes composed of a
mixture of ASLR-enabled and disabled images, which simply means that some images
will get rebased and others will not. Apart from legacy libraries which were
compiled before ASLR existed, a library might not support ASLR because parts of a program use hard-coded jump addresses within that library.
Bypassing ASLR appears to be difficult in practice, with no currently-known
generic attack.
The work by Shacham et al.~\cite{aslr_effectiveness} relies on brute force, which only works if the
vulnerable application does not crash when an access violation occurs.
\textit{Partial overwrites}~\cite{phrack_bypass_pax_aslr} overwrite only the
last two bytes of an address on the stack. Because only the first two bytes get randomized, this
attack does not require knowledge of the randomness introduced by ASLR. This gives an attacker a range of at most 4096 bytes of instructions.
Durden presents information leaks as a way of gathering information about
the memory layout of an ASLR-protected
application~\cite{phrack_bypass_pax_aslr}. Hund et al.~\cite{aslr_sidechannel}
propose a timing-based side channel attack that can
break kernel space ASLR within minutes, given that an attacker knows the
hardware of the attacked system. However, most current exploits do not have to use such
techniques, and instead can rely on the presence of some non-ASLR images on the
target platform. Such images, however, are still very common on current systems,
which is why many attacks still succeed.

\subsection{Attacker Model}
%
We assume a relatively strong attacker, who is able to bypass DEP, ASLR, and other mitigation techniques which are currently part of Windows. This is, for a determined attacker, a realistic assumption. We
even go one step further and allow for pure RoP and JoP attacks, which do not
have to call \texttt{VirtualProtect} or \texttt{VirtualAlloc} but instead call
the malicious code directly.
Such pure code-reuse attacks are still rarely found in the wild, but we expect them to
increase due to the work that is being done on detecting two-staged attacks.
One known example is a pure RoP attack on Adobe Reader~\cite{reader_666}.

\section{AntiCRA}
\label{Sec:AntiCRA}
When designing AntiCRA, we analyzed RoP and JoP exploits and their underlying
principles. We found that the exploits share properties which are unusual and
typically not present in a normal program's execution. Based on these
observations, we built a heuristic which monitors the following two properties:

\subsubsection*{Indirect Branches}
Code-reuse attacks consist of gadgets which all end in an indirect branch. We
analyzed benchmarks as well as real-world applications like Adobe Reader, VLC,
Microsoft Office, Open Office (in total 34 programs; the complete list is available on the project's website) and found that a very high number of consecutive indirect branches is rather unusual. The highest number of subsequent indirect branches we found during our experiments was 47 (in Microsoft Word), but only 8 of the 34 programs execute 15 or more subsequent indirect branches.



\subsubsection*{Average Length of Basic Blocks}
To reduce side-effects on other registers, the stack, or flags, exploit
developers try to use gadgets that are as short as possible. Therefore, at least
for contemporary approaches, gadgets can be considered basic blocks with very
few instructions. As with indirect branches, we analyzed program behaviour of
legitimate programs and found that the average number of instructions over a
sliding window of 10 basic blocks did not drop below 2.33. We also found an
interesting correlation between this and the previous property: the more consecutive
indirect branches, the longer the corresponding basic blocks. We make use of
this knowledge in the next paragraph, when we try to find default parameters
which work for a wide set of applications.

As previously mentioned, since programs can exhibit varying characteristics regarding these two properties, \toolname first runs in learning mode. This requires nothing from the user but simply using the program she wants to protect as usual, while in the background, \toolname observes the program flow and determines appropriate thresholds for these two properties. This, of course, leads only to limited coverage, however for our approach high coverage is not required. Exploiting a buffer overflow requires some sort of input, generally provided by the attacker as a file that has to be opened by the victim and is then processed by the vulnerable program. Thus, a user working with the program covers the important cases which lead to exploitation. For our sample set we chose applications, which are commonly used in corporate and personal environments. As expressed earlier, we recommend setting individual thresholds for different programs, but at the same time we were wondering whether it is possible to provide default values which cover as many programs as possible. After analyzing our test set of benign applications, by running the learning mode and using the programs in our sample set (e.g., opening various media files using VLC, opening various PDF files with Adobe Reader, working with Microsoft Word, etc.) we set the following thresholds: 35 subsequent indirect branches and an average basic block length of 2.25 or lower; as described earlier, we found a correlation that larger numbers of subsequent basic blocks also means longer basic blocks. Therefore we added another threshold; 36 till 50 subsequent indirect branches and an average basic block length of 4 or lower. AntiCRA signals an exploitation attempt if one of the two bounds is violated or if, at any point, more than 50 subsequent indirect branches are executed. While our sample set of benign applications may not be large enough to make a claim, that theses suggested thresholds hold for all programs, they do hold for all programs in our set, which includes some of the most exploited applications. Therefore, they serve as an excellent starting point for fine-tuning, should it be required. Since we included many programs that are often found and exploited in business environments (e.g., Word, Excel, Adobe Reader), \toolname can be deployed immediately without the need to fine-tune thresholds.

To increase performance and make the algorithm less prone to false
positives, calculating averages starts only after we have collected 15
basic-block lengths, i.e., the first computed average is available
only after 15 subsequent indirect branches. This prevents false alarms based on
short sequences of short basic blocks, whose sample size is otherwise not
significant enough.
Figure~\ref{fig:scatterplot} (in Section~\ref{Sec:Eval}) shows how the two
thresholds form a (shaded) area in a two-dimensional plain. If an execution
falls into the shaded area then AntiCRA will signal it as malicious. The figure
also summarizes the results of our empirical evaluation, and will be explained
in more detail later.

\subsection{Impact on Current and Future Exploits}
For a code-reuse attack to circumvent AntiCRA, it must not use
more than 34 / 49 consecutive indirect branches. If this is possible at all
depends on the availability of gadgets, which varies between programs based on
what libraries are loaded and whether or not ASLR is being employed.
Furthermore, the average number of instructions in the gadgets used must never
fall below 2.25 / 3.5. Combined, these restrictions make it very difficult for an attacker to create a pure RoP or JoP payload.
Attackers could attempt to raise the average number of instructions per
gadget by inserting longer gadgets. But longer gadgets usually have unwanted
side-effects, like manipulating other registers that hold important data, or the
stack, or modifying flags.
Furthermore, since the total number of gadgets is limited to 34 / 49, inserting long
gadgets whose side effects are irrelevant just for the sake of increasing the
average wastes precious slots for useful gadgets. To bypass AntiCRA, an attacker would have to try and insert direct branches, but, due to limited availability and
side-effects, this is anything but trivial. In particular, we know of no
gadget compiler that would support direct branches at this point. Depending on
the program it might still be possible, but, as previously mentioned, our goal is to break
current exploits and make the development of new code-reuse exploits
significantly more difficult, which AntiCRA certainly achieves.
Long NOP gadgets, as proposed by Davi et al.~\cite{stitching_gadgets}, could potentially be used to artificially increase the average basic block length, however, it takes five gadgets to restore the stack and registers to their original form. Therefore, precious space has to be wasted and such an attempt will most likely exceed any sensible threshold. Furthermore, the authors state that finding such a gadget was "a non-trivial task that required painstaking analyses and a stroke of luck", so for some programs this technique might not be possible at all.

\subsection{Limitations}
Due to its heuristic nature, false positives as well as false negatives are
possible. As we show in this work, however, in practice the heuristic seems
effective enough to go without any false decisions, at least in our benchmark
set. Furthermore, under circumstances very favorable to an attacker it
might be possible to create a two-staged exploit that disables DEP using fewer
than 15 gadgets and then runs a regular payload. This would not be detected by
AntiCRA and motivates the need for reliably non-executable data sections, which
we enforce using DEP+ (Section~\ref{Sec:DEPPlus}).

\section{DEP+}
\label{Sec:DEPPlus}
DEP+ is based on the same concept as DEP, i.e., the premise
that data should not be executable. DEP+ thus monitors the loading and
unloading of images and creates a virtual memory map based on this information.
All virtual memory space where no image is mapped is considered to hold
potentially malicious data, since Windows can allocate stacks or heaps in these
areas. To enforce that the instruction register EIP never points outside an
image DEP+ checks the register value after each indirect branch, i.e., after
each return, indirect call, and indirect jump. Opposed to DEP, DEP+ cannot be
disabled through API calls such as \texttt{VirtualProtect}.

\subsection{Implementation Details}
PIN's \verb IMG_AddInstrumentFunction  as well as \verb IMG_AddUnloadFunction  are used to monitor the loading and unloading of images.
When an image is loaded, DEP+ stores its start and end address in an array of
structs; if the same image is unloaded at runtime, it is removed from the
array. The data structure
results in a virtual-memory map that distinguishes only between images and
non-images, i.e., code and data regions.
DEP+ treats the latter as space for potentially malicious data, hence does not
allow EIP to point into it. To do so, DEP+ checks if the last instruction of
a basic block is an indirect branch, and, if so, it checks if the target address
of the branch points inside any of the data regions.

Some programs load 30 or more libraries, which means that processes can
have an equally high number of code regions. As we found, checking each of those
regions after each indirect branch can incur a significant performance penalty.
To increase performance, we thus make use of the fact that Windows' memory
management is relatively deterministic. Images, in general, tend to be loaded at very high addresses,
around \texttt{0x60000000} and higher, while stacks and heaps are at low
addresses and new ones are allocated towards increasingly higher addresses.
Depending on the memory usage of a process, it is generally valid to assume that stacks and heaps reside below
most images.

DEP+ makes use of this knowledge by not checking all regions when checking EIP
after an indirect branch. 
DEP+ monitors a program's memory usage to dynamically increase or decrease the
number of regions that need to be taken into account.  We implement this by 
probing memory usage every 10th time a function that allocates or de-allocates
memory is called and multiply the reported usage by 1.3 to have a large enough
safety margin. This is, of course, a heuristic, which trades security for performance,
but as our evaluation in Section~\ref{Sec:Eval} shows, the heuristic helps
DEP+ to bring the checks down to a minimum while still recognizing all tested
attacks. Furthermore, our experiments using benchmarks and real-world programs
have shown that memory allocations are done in many small steps, hence probing
memory usage in short intervals and adding a safety margin of 30\% has never
failed to correctly detect the necessary number of regions which have to be
checked. For the heuristic to fail and be exploitable it would take a
memory allocation of about 30\% of the current memory usage, a vulnerable function which uses this
memory and an instruction which redirects program flow into this memory before
our algorithm checks memory usage again. Since an attacker has no influence on
any of these preconditions, we accept the risk that our heuristic might fail
under rare conditions. Reliably exploiting such circumstances in a
multi-threaded program would be even more difficult, due to their highly non-deterministic nature.

\subsection{Comparison to DEP}
The original shortcomings of DEP are that it may not be enabled at all, or
that it can be bypassed by both pure code-reuse attacks and by code-reuse attacks that
invoke \texttt{VirtualProtect} etc.\ to disable DEP. DEP+ improves over DEP in that
it prevents the execution of injected code by enforcing non-executable data
regions \emph{even} for processes that run with regular DEP disabled.
In particular, DEP+ cannot bypassed by calls to \texttt{VirtualProtect} and its
siblings, as such calls have no effect on DEP+. A similar result could be
achieved by hooking said functions and simply not executing them. However,
userland hooks can be bypassed easily~\cite{phrack_bypass_hooks} and kernel
hooks require administrator privileges, hence make deploying our solution more
complicated. Furthermore, as Section~\ref{Sec:Eval} shows, the overhead
introduced by DEP+ is negligible.

\subsection{Limitations}
Processes which rely on the ability to execute code from outside images, e.g.,
processes which generate code at runtime or incorporate self-modifying code, are
not compatible with DEP+. Such a process is not compatible with DEP
either, unless it uses the \texttt{VirtualProtect} API etc.\ to disable
DEP for memory regions with generated code. Since it is difficult to detect
whether a call to the API usually abused to bypass DEP by an attacker is legitimate, i.e.,
 originating from the program itself, we decided against supporting such calls. This results
in a strong increase in security, at the drawback of slightly reduced
compatibility with mostly older software.

Like DEP, DEP+ cannot detect and thus not prevent the exploitation of the
vulnerability itself, e.g., the overwriting of data on the stack due to a buffer
overflow. Therefore, non-control data attacks~\cite{non_control_data_attacks}
or information leakages are still possible. Furthermore, DEP+ does not prevent
\emph{pure} code-reuse attacks, motivating the need for AntiCRA
(Section~\ref{Sec:AntiCRA}).

\section{Evaluation}
\label{Sec:Eval}
Our implementation is highly modular, so that one may deploy AntiCRA or DEP+
independently as well as in combination. Running both of them, however, strongly
increases security, in a similar fashion as running with DEP and ASLR.

In this chapter we evaluate AntiCRA and DEP+ by addressing the following
research questions:


\begin{compactitem}
	\item RQ1: How effectively does AntiCRA detect pure code-reuse payloads?
	\item RQ2: How effectively does AntiCRA detect two-staged RoP payloads?
	\item RQ3: How effectively does DEP+ detect code-injection attacks?
	\item RQ4: What is the performance overhead of AntiCRA and DEP+?
\end{compactitem}

\subsection{Evaluation of AntiCRA (RQ1/RQ2)}
\label{Sec:EvalAntiCRA}
For evaluating RQ1 we looked at pure code-reuse attacks, however, at this point such payloads are only rarely found in the wild and are mostly used in
academia as proof of concept. The only real-world pure code-reuse exploit we
found is a RoP exploit for Adobe Reader. Since neither the exploit's source code, nor an
infected file are publicly available, our conclusion is based on an analysis by
Li and Szor~\cite{reader_666}. Analyzing the exploit's source code reveals that the
address \texttt{0x6acc1049} is repeated 9,344 times; the instruction at that address is a simple \texttt{ret}. This equals to over 9,000
indirect branches in a row, which would, of course, be detected by AntiCRA.

The likely reason for why pure RoP and JoP payloads still seem to be rare in
practice is that two-staged payloads (which aim to disable DEP through RoP/JoP)
are simpler to construct and are sufficient in many cases. Such payloads can be
mitigated by DEP+, but nevertheless we were interested in evaluating RQ2, i.e.,
to what extent AntiCRA alone, without DEP+, can be used to mitigate such attacks
as well.

We analyzed 11 real-world exploits in total. To operate on an unbiased test set, we analyzed the 10 most recent
exploits from \url{http://www.toexploit.com/} which claim to bypass ASLR and also added the previously mentioned pure RoP exploit. Figure~\ref{fig:scatterplot} shows the results of our analysis, i.e., the number of consecutive indirect branches and the average basic block length for each exploit and also for legitimate programs. As the numbers indicate, legitimate programs rarely have more than 15 consecutive indirect branches and their average basic block length is higher than that of exploits. This confirms that our generalized threshholds, which work for a wide variety of programs, are well-suited to detect attacks.

AntiCRA detects 10 out of the 11 exploits in our sample set. In five cases this
is due to the number of indirect branches in a row. Three exploits are detected because they use very
short gadgets, which mostly only execute one instruction and then transfer
program execution to the next gadget. Two exploits trigger both mechanisms, since they use more than 35 indirect
branches in a row and also very short gadgets.

One exploit cannot be detected by AntiCRA. This is because it requires
only 13 gadgets to prepare the stack for calling \texttt{VirtualProtect}. This is not
enough to trigger the indirect-branch check. The average
length of the basic blocks is 2.2, which would trigger an alarm. However, as
explained in Section~\ref{Sec:AntiCRA}, we only trigger inspections after a
total of 15 indirect branches in a row.

\begin{figure}[h]
\centering
\begin{tikzpicture}
	\begin{axis}[%
	scatter/classes={%
		a={mark=square*,blue},%
		b={mark=triangle*,red},%
		c={mark=o,draw=black}},
		xlabel=No. of indirect branches in a row,
		ylabel=Lowest average of BBs,
		legend entries={SPEC, Exploits, Applications},
    	legend style={nodes=right},
    	legend pos=south east
		]
		
\draw[red] [thick] (120,-15) -- (120,225);
\draw[red] [thick] (120,225) -- (320,225);
\draw[red] [thick] (320,225) -- (320,400);
\draw[red] [thick] (320,400) -- (500,400);

\draw[black!50,dotted] [thin] (-15,225) -- (320,225);

\path[fill=red!20] (120,-15) -- (120,225) -- (320,225) -- (320,400)
	-- (500,400) -- (500,-15);

\node[font=\tiny] at (50,230) {Threshold BB length};
\node[font=\tiny] at (240,330) {Threshold ind. branches};
		
\addplot[scatter,only marks,%
		scatter src=explicit symbolic]%
	table[meta=label] {
x      		y		label
50			2.5		b		
20			1.9		b
16			2		b
17			2		b
13			2.2		b
43			2		b
49			2.5		b
46			2.5		b
48			2.5		b
50			2		b
50			1.5		b
4			0		a
3			0		a
6			0		a
3			0		a
3			0 		a
4			0		a
5			0		a
7			0		a
8			0		a
6			0 		a
3			0		a
3			0		a
31			4		a
4			0		a
15			3.91	a
3			0		a
9			0		a
17			4		a
14			0		c
14			0		c
29			2.33	c
8			0		c
9			0		c
40			4.97	c
13			0		c
47			4.14	c
28			4.1		c
25			4.14	c
7			0		c
7			0		c
7			0		c
11			0		c
5			0		c
};
\end{axis}
\end{tikzpicture}
\caption{Analysis of the number of indirect branches in a row and the lowest
average basic block length of our test set} 
\label{fig:scatterplot}
\end{figure}
%

It is important to point out that the two-staged exploit AntiCRA misses is detected by DEP+. AntiCRA is primarily designed to catch pure RoP and JoP attacks, not necessarily the
two-staged attacks like the ones examined in the evaluation. It is also important to keep in mind that the thresholds can and should be adjusted for each program and that this section evaluates how well our generalized thresholds hold. Despite this, it
still detects 10 out of 11 exploits.  Because of these results and our
analysis of the pure RoP exploit for Adobe Reader, we are very
confident that exploits which rely solely on RoP or JoP can be detected by AntiCRA.

\subsection{Evaluation of DEP+ (RQ3)}
To test DEP+, we wrote a small vulnerable application, which uses an unbounded \texttt{strcpy} and was compiled with the \verb NX_COMPAT, and a simple exploit.
Since all code injection attacks store the injected code inside a buffer which,
by definition, cannot be in an image, the program that contains the
vulnerability is of little consequence. The only differences between our
vulnerable application and a real application are mitigation techniques which might be in
place, but which are irrelevant to us, since we assume an attacker is able to
bypass them, and how program flow is transferred to the injected code, which is
irrelevant for our evaluation as well.
Ultimately, all code injection attacks end up calling their injected code, and
this is where DEP+ detects them. Therefore, evaluating DEP+ with this
self-written program poses no real threat to the validity of this experiment.
As expected, DEP+ correctly detects that the target address of the \texttt{ret}
instruction at the end of our vulnerable function is not
in an image, before the instruction is actually executed. Therefore, it can
terminate the program and mitigate an attack, which would have led to arbitrary
code execution.
As for the real world exploits, DEP+ detects each one except for the pure RoP exploit for Adobe Reader,
as all the others eventually do execute code from memory outside of images.

\subsection{Performance (RQ4)}
We evaluated the performance of \toolname using the C and C++ benchmarks in the
SPEC CPU2006 benchmark suite. 
Note that those are really worst-case benchmarks that exercise the dynamic
analysis heavily. Any interactive or network-based application would show a
significantly lower overhead. We measured five different runtimes for each
benchmark:
\begin{compactitem}
	\item The native runtime, i.e., without PIN.
	\item The runtime with PIN attached, but without instrumentation, to get the
	basic overhead PIN introduces.
	\item The runtime with AntiCRA.
	\item The runtime with DEP+.
	\item The runtime with AntiCRA and DEP+.
\end{compactitem}
Benchmarks
were run on Windows 7 SP1 with an Intel Core 2 Duo T9400 clocked at 2.53
GHz and 4 GB RAM using the reference workload.

Figure~\ref{fig:perf} summarizes the results of our performance benchmarks.
Running a program under PIN but without any instrumentation introduces an
average overhead\footnote{Average overheads were computed using the geometric
mean, which is considered best practice for reporting normalized values such as percentages of overhead~\cite{Fleming:1986:LSC:5666.5673}} of 1.36x, i.e., programs take, on average, 36\% more time to finish, ranging from 1.002x (470.lbm)
to 2.24x (464.h264ref). Programs protected by AntiCRA run, on
average, with a total overhead 2.2x. With DEP+
enabled as well, \toolname introduces an average overhead of 2.39x, which is
comparable to similar tools such as ROPdefender~\cite{ropdefender}, which gives
weaker guarantees. Compared to CFI approaches~\cite{cfi_for_cots, kbouncer, ccfir} \toolname has a considerably higher overhead, however, it monitors a process throughout its whole lifetime and not just at potentially dangerous points. Thus, it can determine more accurate if a RoP attack is being carried out.

\begin{figure*}
\begin{tikzpicture}
\pgfplotstableread{ 
Label   First   Second  Third	Fourth
400		570     401     820		367
401     689     306      69		 46
403     491     420     378		102
429     370      18       5		  3
433		775		 34		582		 10
444		718		 7		  8		  2
445     628     422     448		 66
447		596	    182		371		 31
450		520		 50		460		  8
453		321		161	   2069		103
456		1295	 18		129		 53
458		737	    472		748		350
464		999	   1236    2702		225
470		798		  2		  5		  3
471		406		 90		941		 83
473		477		153		168		 19
482		862		168  	607	   1466
483		346		332		959		183

}\datatable

\begin{axis}[
    ybar stacked,   
    ylabel=Runtime (s),
    ymin=0,         
    height=5.5cm,
    width=0.95\linewidth,
    bar width=10pt,
    xtick=data,     
    xticklabels from table={\datatable}{Label},  
    x tick label style={rotate=45, anchor=north east},
    xlabel=SPEC CPU2006 Benchmark,
    legend entries={Native, PIN only, PIN \& AntiCRA, PIN \& AntiCRA \&
    DEP+},
    legend style={nodes=right},
    legend pos=north west
]
\addplot [fill=white] table [y=First, x expr=\coordindex]
{\datatable};
\addplot [fill=black!10]table [y=Second, x expr=\coordindex]
{\datatable};
\addplot [fill=black!50!blue!30] table [y=Third, x expr=\coordindex]
{\datatable};
\addplot [fill=black!50!blue!50] table [y=Fourth, x expr=\coordindex]
{\datatable};
\end{axis}
\end{tikzpicture}
\caption{Performance of \toolname}
\label{fig:perf}
\end{figure*}

While overheads in the order of two-fold might sound unacceptable, those
overheads should really only be expected in worst-case situations.
Thus, while performance benchmarks such as SPEC CPU2006 are advantageous in
producing reproducable results, the results that they do produce do not reflect
reality very much.
What ultimately counts is the performance on real-world applications.
Their performance can, however, often hardly be measured systematically, which is why
we only report qualitative results on some of the applications in our sample set.
As a general observation we can say that in all cases the GUI had some slight
input lag $<$ 1 second when opening a menu for the first time, however,
afterwards they opened in an instant. File transfers with Filezilla were no
slower than without our tool. VLC plays h.264 encoded HD videos without any
jitter. Adobe Reader renders pages without any noticeable lag. Typing in
Microsoft Word has no input lag. We want to emphasize that \toolname is not intended to be used with all applications at all times. Instead, our recommended usage is to enable it only for either very critical systems, or for an application which has a vulnerability that's being actively exploited and no vendor patch has been released yet. Under such circumstances the overhead is, in our opinion, acceptable.

\section{Related Work}
\label{Sec:Related}

\textbf{TRUSS}~\cite{truss} and \textbf{ROPdefender}~\cite{ropdefender} store copies of return addresses using a runtime shadow stack. When a function is called, a copy of
the pushed return address is stored on the shadow stack. Upon returning from a
procedure, the return address on the stack is compared to the one on the shadow
stack. TRUSS is implemented using DynamoRIO, ROPdefender using PIN. DynamoRIO is a
runtime instrumentation tool which works similarly to PIN. TRUSS and ROPdefender
rely on an attacker overwriting the return address on the stack, which is not strictly required. Therefore, they can miss some classes of attacks like JoP.
Furthermore, they assume that function calls are always made through
\texttt{call} and exited via \texttt{ret}. ROPdefender can handle exceptions,
but neither can handle hand-crafted assembly code, which does not necessarily
follow these conventions. The overhead of both tools is similar to ours.

\textbf{kBouncer}~\cite{kbouncer} makes use of the last branch record (LBR)
feature some modern CPUs have. kBouncer assumes that at some point shellcode has
to invoke a system call. When this occurs, the LBR repository is checked for
distinctive properties of ROP-like behaviour, e.g. consecutive indirect jumps
and short basic blocks. The tool has an average overhead of only 1\%,
however, the implementation for Windows 7 is not fully functional, since
Windows 7 does not allow to intercept system calls which is a requirement of
kBouncer. Furthermore, it cannot be deployed on systems whose CPU
doesn't have LBR.

\textbf{ROPecker}~\cite{ropecker} also uses the LBR feature of some modern CPUs.
Like \toolname it checks for consecutive short indirect branches and raises an alert when a certain threshold is undercut. To
increase performance the detection heuristic is only invoked if the branch
target is outside the so-called ``sliding window'' (a collection of pages,
usually 2 or 4, i.e. 8 or 16 kB). Due to these two circumstances, ROPecker has a very low overhead of only 2.6\% for the SPEC CPU2006
benchmark suite. It does, however, miss ROP gadgets which are within the sliding
window and requires a CPU which supports the LBR feature.

\textbf{Control Flow Integrity}~\cite{control_flow_integrity} uses static
analysis of a binary to create a control-flow graph and rewrites the binary to
enforce it does not deviate from the pre-computed paths. The implementation is based on Vulcan, a commercial dynamic instrumentation
tool for x86 binaries. The average overhead is about 16\%. 

\textbf{CCFIR}~\cite{ccfir} enforces control-flow integrity by
ensuring that targets of indirect jumps are legal. Valid targets are identified
ahead of time by statically analysing a given binary. For their analysis to work
properly they require the binary to use ASLR and DEP. CCFIR has a runtime
overhead of about 4\%.

G\"{o}kta\c{s} et al.~\cite{size_matters} have recently shown, that the above mentioned CFI approaches can be bypassed. The inherent problems of these approaches is that there are too few checks, allowing attackers to access too many gadgets. They are further limited by the number of slots in the LBR, which is at most 16. To improve the security of approaches which attempt to detect RoP exploits by measuring similar properties as we do, they suggest making the thresholds dynamic.

\section{Conclusion}
\label{Sec:Conclusion}
In this work we have presented \toolname, a novel tool for the automated
dynamic recognition of buffer-overflow attacks. \toolname is designed to
recognize different classes of code-reuse attacks based on two novel techniques
AntiCRA and DEP+. AntiCRA is a configurable heuristic based on the number of indirect
branches executed in a row as well as on the average basic block length of
executed code. In our experiments using default thresholds which work for a variety of programs,
AntiCRA detects 10 out of 11 of the latest
real-world code-reuse exploits and yields no false alarms on SPEC CPU2006 and
all tested real-world applications, a total of 35 programs. DEP+ executes a non-executable stack through
binary instrumentation and can thus be used to detect exploits based
on two-staged payloads that use a code-reuse attack to disable DEP using the
Windows API. DEP+ successfully detects all two-staged payloads we examined,
again with no false alarms. By combining both techniques, \toolname thus
successfully detects all tested exploits, without false warnings, showing an
average performance overhead of 2.4x for SPEC CPU2006 and real-world
applications showing only an initially noticeable input lag and no stutter.
\toolname runs in user mode, requiring no access to source code, nor debug
symbols or changes to the operating system. It supports multi-threaded
applications. Due to its heuristic nature, \toolname cannot give an absolute
security guarantee. However, the parameters the heuristic is based on should make it very
hard to circumvent the approach in practice. \toolname is thus raising the bar
significantly, without any added cost compared to previous related approaches.

\section*{Acknowledgments}\label{sec:Acknowledgments}

This work was supported by the German Federal Ministry of Education and Research (BMBF) within EC SPRIDE and by the Hessian LOEWE excellence initiative within CASED.

\bibliographystyle{abbrv}
\bibliography{papers}

\begin{thebibliography}{10}

\bibitem{control_flow_integrity}
M.~Abadi, M.~Budiu, U.~Erlingsson, and J.~Ligatti.
\newblock Control-flow integrity principles, implementations, and applications.
\newblock {\em ACM Trans. Inf. Syst. Secur.}, 13(1):4:1--4:40, Nov. 2009.

\bibitem{xp_sp2_dep}
S.~Andersen and V.~Abella.
\newblock Changes to functionality in windows xp service pack 2 - part 3:
  Memory protection technologies, Aug. 2004.

\bibitem{jop2}
T.~Bletsch, X.~Jiang, V.~W. Freeh, and Z.~Liang.
\newblock Jump-oriented programming: a new class of code-reuse attack.
\newblock ASIACCS '11, pages 30--40. ACM, 2011.

\bibitem{bo_top_25}
C.~Bubinas.
\newblock Buffer overflows are the top software security vulnerability of the
  past 25 years, Mar. 2013.

\bibitem{phrack_bypass_hooks}
J.~Butler and Anonymous.
\newblock Bypassing 3rd party windows buffer overflow protection.
\newblock {\em Phrack}, 11, 2004.

\bibitem{jop}
S.~Checkoway, L.~Davi, A.~Dmitrienko, A.-R. Sadeghi, H.~Shacham, and
  M.~Winandy.
\newblock Return-oriented programming without returns.
\newblock CCS '10, pages 559--572. ACM, 2010.

\bibitem{non_control_data_attacks}
S.~Chen, J.~Xu, E.~C. Sezer, P.~Gauriar, and R.~K. Iyer.
\newblock Non-control-data attacks are realistic threats.
\newblock SSYM'05, pages 12--12. USENIX Association, 2005.

\bibitem{ropecker}
Y.~Cheng, Z.~Zhou, M.~Yu, X.~Ding, and R.~H. Deng.
\newblock Ropecker: A generic and practical approach for defending against rop
  attacks.
\newblock 2014.

\bibitem{stitching_gadgets}
L.~Davi, A.-R. Sadeghi, D.~Lehmann, and F.~Monrose.
\newblock Stitching the gadgets: On the ineffectiveness of coarse-grained
  control-flow integrity protection.
\newblock In {\em Proc. of the 23rd USENIX Conf. on Security}, SEC'14, pages
  401--416. USENIX Association, 2014.

\bibitem{ropdefender}
L.~Davi, A.-R. Sadeghi, and M.~Winandy.
\newblock Ropdefender: a detection tool to defend against return-oriented
  programming attacks.
\newblock ASIACCS '11, pages 40--51. ACM, 2011.

\bibitem{phrack_bypass_pax_aslr}
T.~Durden.
\newblock Bypassing pax aslr protection.
\newblock {\em Phrack}, 11, 2002.

\bibitem{Fleming:1986:LSC:5666.5673}
P.~J. Fleming and J.~J. Wallace.
\newblock How not to lie with statistics: the correct way to summarize
  benchmark results.
\newblock {\em Commun. ACM}, 29(3):218--221, Mar. 1986.

\bibitem{size_matters}
E.~G\"{o}kta\c{s}, E.~Athanasopoulos, M.~Polychronakis, H.~Bos, and
  G.~Portokalidis.
\newblock Size does matter: Why using gadget-chain length to prevent code-reuse
  attacks is hard.
\newblock In {\em Proc. of the 23rd USENIX Conf. on Security Symposium},
  SEC'14, pages 417--432. USENIX Association, 2014.

\bibitem{pwn2own}
A.~Gunn.
\newblock Pwn2own 2014: A recap, 2014.

\bibitem{microsoft_aslr}
M.~Howard, M.~Miller, J.~Lambert, and M.~Thomlinson.
\newblock Windows isv software security defenses, Dec. 2010.

\bibitem{aslr_sidechannel}
R.~Hund, C.~Willems, and T.~Holz.
\newblock Practical timing side channel attacks against kernel space aslr.
\newblock SP '13, pages 191--205. IEEE Computer Society, 2013.

\bibitem{reader_666}
X.~Li and P.~Szor.
\newblock Emerging stack pivoting exploits bypass common security, May 2013.

\bibitem{Luk:2005:PBC:1065010.1065034}
C.-K. Luk, R.~Cohn, R.~Muth, H.~Patil, A.~Klauser, G.~Lowney, S.~Wallace, V.~J.
  Reddi, and K.~Hazelwood.
\newblock Pin: building customized program analysis tools with dynamic
  instrumentation.
\newblock PLDI '05, pages 190--200. ACM, 2005.

\bibitem{microsoft_dep}
Microsoft.
\newblock Data execution prevention.

\bibitem{jop_windows}
J.-W. Min, S.-M. Jung, D.-Y. Lee, and T.-M. Chung.
\newblock Jump oriented programming on windows platform (on the x86).
\newblock volume 7335 of {\em Lecture Notes in Computer Science}, pages
  376--390. Springer, 2012.

\bibitem{aleph}
A.~One.
\newblock Smashing the stack for fun and profit.
\newblock {\em Phrack}, 7, 1996.

\bibitem{kbouncer}
V.~Pappas, M.~Polychronakis, and A.~D. Keromytis.
\newblock Transparent rop exploit mitigation using indirect branch tracing.
\newblock SEC'13, pages 447--462. USENIX Association, 2013.

\bibitem{rop}
R.~Roemer, E.~Buchanan, H.~Shacham, and S.~Savage.
\newblock Return-oriented programming: Systems, languages, and applications.
\newblock {\em ACM Trans. Inf. Syst. Secur.}, 15(1):2:1--2:34, Mar. 2012.

\bibitem{rop_org}
H.~Shacham.
\newblock The geometry of innocent flesh on the bone: return-into-libc without
  function calls (on the x86).
\newblock CCS '07, pages 552--561. ACM, 2007.

\bibitem{aslr_effectiveness}
H.~Shacham, M.~Page, B.~Pfaff, E.-J. Goh, N.~Modadugu, and D.~Boneh.
\newblock On the effectiveness of address-space randomization.
\newblock CCS '04, pages 298--307. ACM, 2004.

\bibitem{truss}
S.~Sinnadurai, Q.~Zhao, and W.~fai Wong.
\newblock Transparent runtime shadow stack: Protection against malicious return
  address modifications.

\bibitem{ccfir}
C.~Zhang, T.~Wei, Z.~Chen, L.~Duan, L.~Szekeres, S.~McCamant, D.~Song, and
  W.~Zou.
\newblock Practical control flow integrity and randomization for binary
  executables.
\newblock SP '13, pages 559--573. IEEE Computer Society, 2013.

\bibitem{cfi_for_cots}
M.~Zhang and R.~Sekar.
\newblock Control flow integrity for cots binaries.
\newblock In {\em Proc. of the 22Nd USENIX Conf. on Security}, SEC'13, pages
  337--352. USENIX Association, 2013.

\end{thebibliography}
%


%
%

\end{document}